\documentclass[english]{amsart}
\usepackage[T1]{fontenc}
\usepackage[latin9]{inputenc}
\usepackage{geometry}
\usepackage{amsthm}
\usepackage{amssymb}
\usepackage{graphicx}

\makeatletter
\numberwithin{equation}{section}
\numberwithin{figure}{section}

\makeatother

\usepackage{babel}
\begin{document}

\title{Relative equilibria and relative periodic solutions in systems with
time-delay and $S^{1}$ symmetry}

\author{Serhiy Yanchuk$^{1}$ and Jan Sieber$^{2}$ }

\address{$^{1}$Institute of Mathematics, Humboldt University of Berlin, Unter
den Linden 6, 10099 Berlin,\\
Email: yanchuk@math.hu-berlin.de}

\address{$^{2}$University of Exeter, UK}

\maketitle

\begin{abstract}
We study properties of basic solutions in systems with dime delays and $S^1$-symmetry. 
Such basic solutions are relative equilibria (CW solutions) and relative periodic 
solutions (MW solutions). It follows from the previous theory that the number of CW
solutions grows generically linearly with time delay $\tau$. Here we show, in particular, 
that the number of relative periodic solutions grows generically as $\tau^2$ when delay
increases. Thus, in such systems, the relative periodic solutions are more abundant than
 relative equilibria. The results are directly applicable to, e.g., Lang-Kobayashi model for
 the lasers with delayed feedback. We also study stability properties of the solutions
for large delays. 
\end{abstract}

\section{Introduction}

We consider delay differential equations 
\begin{equation}
x'(t)=f(x(t),e^{A\varphi}x(t-\tau))\label{eq:main}
\end{equation}
which are $S^{1}$-equivariant 
\begin{equation}
f(e^{A\theta}x,e^{A\theta}y)=e^{A\theta}f(x,y).\label{eq:symm}
\end{equation}
Here $f:\,\mathbb{R}^{n}\times\mathbb{R}^{n}\to\mathbb{R}^{n}$ is
smooth, $A$ is $n\times n$ matrix satisfying $A^{T}=-A$, $\theta\in S^{1}$,
and $e^{A\theta}$, $e^{A2\pi}=I$ is a representation of $S^{1}$
symmetry group in $\mathbb{R}^{n}$. 

Systems of the form (\ref{eq:main}) include many practically important
models. For example, the Lang-Kobayashi (LK) system describing the
dynamics of a semiconductor laser with delayed feedback. In dimensionless
coordinates, LK system has the form \cite{Lang1980,Yanchuk2010a,Heil2001}
\begin{equation}
\begin{array}{c}
E'(t)=(1+i\alpha)NE(t)+\eta e^{i\varphi}E(t-\tau),\\
N'(t)=\varepsilon\left[J+N+(2N+1)|E(t)|^{2}\right],
\end{array}\label{eq:LK}
\end{equation}
where $N\in\mathbb{R},$ $E\in\mathbb{C}$ are variables and $\varepsilon,J,\alpha,\eta,$
and $\varphi$ are real parameters. System (\ref{eq:LK}) has the
form (\ref{eq:main}) and satisfies the equivariance condition (\ref{eq:symm})
with 
\[
e^{A\theta}=\left[\begin{array}{ccc}
\cos\theta & -\sin\theta & 0\\
\sin\theta & \cos\theta & 0\\
0 & 0 & 0
\end{array}\right],\quad A=\left[\begin{array}{ccc}
0 & -1 & 0\\
1 & 0 & 0\\
0 & 0 & 0
\end{array}\right],\quad x=\left[\begin{array}{c}
\Re(E)\\
\Im(E)\\
N
\end{array}\right].
\]
Another example is the limit cycle oscillator with delayed feedback
\[
z'(t)=\left(\alpha+i\beta+\gamma|z|^{2}\right)z+\eta e^{i\varphi}z(t-\tau),\quad z\in\mathbb{C}.
\]
A large class of models of the form (\ref{eq:main}) are coupled lasers
and coupled limit cycle oscillators. 

Note that the additional parameter $\varphi$ in (\ref{eq:main})
appears naturally in all above mentioned models. Moreover, as it follows
from this study, it is a natural parameter in such symmetric systems.

Our results can be extended to systems with multiple time delays 

\[
x'(t)=f(x(t),e^{A_{1}\varphi_{1}}x(t-\tau_{1}),\dots,e^{A_{k}\varphi_{k}}x(t-\tau_{k}))
\]
in a straightforward way. In order to avoid non-essential technicalities,
we limit our presentation to one delay.

\section{Relative equilibria and relative periodic solutions\label{sec:CW-MW}}

Due to the symmetry, system (\ref{eq:main}) generically possesses
relative equilibria of the form 
\begin{equation}
x(t)=e^{A\omega t}x_{0},\label{eq:eq}
\end{equation}
which are periodic solutions with period $2\pi/\omega$. We call these
solutions \emph{continuous waves} (CW). The trajectories of CW coincide
with the group orbit $e^{A\theta}x_{0}$. Substituting (\ref{eq:eq})
into (\ref{eq:main}), CWs can be obtained from the following algebraic
system 
\begin{equation}
f\left(x_{0},e^{A(\varphi-\omega\tau)}x_{0}\right)-A\omega x_{0}=0.\label{eq:eqforeq}
\end{equation}
Since $x_{0}$ is defined up to the symmetry shift, system (\ref{eq:eqforeq})
should be additionally augmented by a condition on $x_{0}$, e.g.

\begin{equation}
b^{T}x_{0}=0\label{eq:condx0}
\end{equation}
with some $b\in\mathbb{R}^{n}$ such that $b\not\in\ker A$. The system
of $n+1$ equations (\ref{eq:eqforeq}) and (\ref{eq:condx0}) determines
$\omega$ and $x_{0}$. 

For example, for the LK system, the CW solutions have the form $E(t)=E_{0}e^{i\omega t},$
$N(t)=N_{0}$, where $E_{0}$ can be chosen real. 

\emph{Modulated waves} (MW) are solutions of the form 
\begin{equation}
x(t)=e^{A\omega t}a(\beta t),\label{eq:per}
\end{equation}
where $a(\cdot)$ is a $2\pi$-periodic function. Substituting (\ref{eq:per})
into (\ref{eq:main}), the equation for the function $a(\cdot)$ has
the form 
\begin{equation}
\beta\frac{da(y)}{dy}=-A\omega a(y)+f\left(a(y),e^{A\left(\varphi-\omega\tau\right)}a(y-\beta\tau)\right),\label{eq:eqper}
\end{equation}
\begin{equation}
a(0)=a(2\pi).\label{eq:eqperbc}
\end{equation}
(\ref{eq:eqper}) -- (\ref{eq:eqperbc}) is an autonomous periodic
boundary value problem, therefore, MW are expected to appear generically
in system (\ref{eq:main}) and the relative frequencies $\beta$ and
$\omega$ are smooth functions of the system parameters.

\section{Properties of continuous and modulated waves}

In this section, we consider basic properties of CWs and MWs. The
main emphasis will be made on their dependence on parameters: delay
$\tau$ and feedback phase $\varphi$.

\subsection{Continuous waves\label{sub:CW}}

The following result summarizes basic properties of CWs. In particular,
it shows that any CW, which exists for some parameter values $(\varphi,\tau)$,
exists also for $\tau=0$. With the increasing of $\tau$, the number
of coexisting CWs grows linearly and their stability approaches some
asymptotic value determined by a pseudo-continuous spectrum of eigenvalues. 

\textbf{Theorem 1. }\\
\emph{1. }\textbf{\emph{{[}Equation for CW{]}}}\emph{ CWs of system
(\ref{eq:main}) can be obtained as solutions of (\ref{eq:eqforeq})--(\ref{eq:condx0}).}\\
\emph{2. }\textbf{\emph{{[}Reapearrance{]}}}\emph{ If a CW $e^{A\omega t}x_{0}$
exists for parameter values $(\tau_{0},\varphi_{0})$, then the same
solution exists for all parameters $(\tau,\varphi)$ satisfying }
\begin{equation}
\varphi-\omega\tau=\varphi_{0}-\omega\tau_{0}\,\,\mathrm{mod}\,\,2\pi.\label{eq:phitaucw}
\end{equation}
\emph{3. }\textbf{\emph{{[}Primary set of CWs{]}}}\emph{ The set of
all CWs existing at $\tau=0$ 
\begin{equation}
e^{A\Omega(\psi)t}X_{0}(\psi),\quad\psi\in S^{1}\label{eq:PRIMSET}
\end{equation}
is called primary set of CWs. $\Omega(\psi)$ and $X_{0}(\psi)$ can
be composed of several branches $\Omega^{(k)}(\psi)$ and $X_{0}^{(k)}(\psi),\dots N$,
where the functions $\Omega^{(k)}(\psi)$ and $X_{0}^{(k)}(\psi)$
are smooth with the exception of maybe finite number of points. The
primary set contains all CWs, which exist in (\ref{eq:main}) for
any $\varphi$ and $\tau\ge0$. }\\
\emph{4. }\textbf{\emph{{[}Coexistence of CWs versus delay{]}}}\emph{
With the increasing of $\tau$, the number of coexisting CWs grows
linearly with $\tau$. More specifically, given any continuous subset
of the primary set of CWs (\ref{eq:PRIMSET}) for $\psi\in[\psi_{1},\psi_{2}]\in S^{1}$
such that $\Omega(\psi)$ is monotone on }$[\psi_{1},\psi_{2}]$,\emph{
there appear 
\begin{equation}
N=\left[K\tau\right]_{\mathrm{int}}\pm1\label{eq:Ktau}
\end{equation}
 CWs for any $(\tau,\varphi)$ from this family. Here $\left[\cdot\right]_{\mathrm{int}}$
is the integer value and the coefficient $K$ is given by 
\begin{equation}
K=\frac{1}{2\pi}\left(\Omega_{\max}-\Omega_{\min}\right),\label{eq:K}
\end{equation}
\[
\Omega_{\max}=\max_{\psi\in[\psi_{1},\psi_{2}]}\Omega(\psi),\quad\Omega_{\max}=\min_{\psi\in[\psi_{1},\psi_{2}]}\Omega(\psi).
\]
If $\Omega(\psi)$ is non-monotone along the primary set, then the
same relation (\ref{eq:Ktau}) holds true, where $K$ is the total
variation of }$\Omega(\psi)$.\emph{}\\
\emph{5. }\textbf{\emph{{[}Stability{]}}}\emph{ Stability of CW $x_{0}e^{iA\omega t}$
at $(\tau,\varphi)$ is determined by the roots of the following characteristic
equation 
\begin{equation}
\det\left[\lambda I-M_{1}+A\omega-M_{2}e^{-\lambda\tau}\right]=0,\label{eq:CWcheq}
\end{equation}
where $M_{1}=D_{1}f(x_{0},e^{A\psi}x_{0})$, $M_{2}=D_{2}f(x_{0},e^{A\psi}x_{0})e^{A\psi}$,
and $\psi=\varphi-\omega\tau$. In particular, if all eigenvalues
except the trivial one $\lambda_{0}=0$ have negative real parts,
then the CW is asymptotically exponentially stable. }\\
\emph{6. }\textbf{\emph{{[}Asymptotic stability for large delay{]}}}\emph{
For large delays $\tau$ the stability of CW $x_{0}e^{iA\omega t}$
is asymptotically determined by the following two spectra:}

\emph{i) Asymptotic continuous spectrum
\[
\gamma_{j}(\chi)=-\frac{1}{2}\ln|Y_{j}(\chi)|
\]
where $Y_{j}(\chi)$, $j=1,\dots,\tilde{n}$ are roots of the polynomial
\[
\det\left[i\chi I-M_{1}+A\omega-M_{2}Y\right]=0
\]
parameterized by $\chi$. }

\emph{ii) Strong point spectrum $\Lambda_{j}$, $j=1,\dots,\tilde{m}$
given by the roots of the polynomial 
\begin{equation}
\det\left[\Lambda I-M_{1}+A\omega\right]=0.\label{eq:SU}
\end{equation}
In particular, under the nongeneracy condition $\mathrm{\ker}M_{2}=\ker M_{2}^{2}$,
the following necessary and sufficient stability conditions hold:}\\
\emph{-- If all roots $\Lambda$ of (\ref{eq:SU}) have negative real
parts and $\gamma_{j}(\chi)<0$ for all $j$ and $\chi\in\mathbb{R}$
except one point where $\chi_{j_{0}}(0)=0$, then the corresponding
CW is asymptotically exponentially stable for all large enough $\tau$.
The point $\chi_{j_{0}}(0)=0$ corresponds to the trivial multiplier
of the periodic solution $x_{0}e^{iA\omega t}$.}\\
\emph{-- Conversely, if there exists either a root $\Lambda$ of (\ref{eq:SU})
with positive real part or $\gamma_{j}(\chi)>0$ for some $\chi$
and $j$, then the corresponding CW is exponentially unstable for
all large enough $\tau$.}

\textbf{Remark.} Statement 6 of theorem 1 implies that the whole primary
family of CWs contains the following open subsets, corresponding to
different asymptotic (for large $\tau$) stability properties: \\
-- strongly unstable CWs, which possess a strong spectrum $\Lambda$
with positive real parts. The largest eigenvalues of such solutions
are close to $\Lambda$ for large $\tau$.\\
-- weakly unstable CWs, which possess unstable asymptotic continuous
spectrum \emph{$\gamma_{j}(\chi)>0$} for some $\chi$ and $j$ and
all $\Lambda$ with negative real parts. The largest eigenvalues of
such solutions are located close to $\lambda=\gamma_{j}(\chi)/\tau+i\chi$
for large $\tau$;\\
-- asymptotically stable CWs.

\subsection{Modulated waves}

The following theorem presents basic properties of MWs. 

\textbf{Theorem 2. }\\
\emph{1. }\textbf{\emph{{[}Equation for MW{]}}}\emph{ MWs of system
(\ref{eq:main}) can be obtained as solutions of (\ref{eq:eqper}).
}\\
\emph{2. }\textbf{\emph{{[}Reappearance{]}}}\emph{ If a MW }$e^{A\omega t}a(\beta t)$\emph{
exists for the parameter values $(\tau_{0},\varphi_{0})$, then the
same solution exists for $(\tau,\varphi)$ parameters, which satisfy
\begin{equation}
\tau=\tau_{0}+\frac{2\pi}{\beta}k,\label{eq:taurule}
\end{equation}
\begin{equation}
\varphi=\varphi_{0}+\frac{2\pi\omega}{\beta}k\,\,\mathrm{mod}\,2\pi,\label{eq:phirule}
\end{equation}
where $k\in\mathbb{Z}$.}\\
\emph{ 3. }\textbf{\emph{{[}Primary set of MWs{]} }}\emph{The set
of all MWs can be parametrized by two parameters $(\tau,\varphi)$
with }$\tau<2\pi/\beta(\tau,\varphi)$.\emph{ The set 
\begin{equation}
e^{A\omega(\tau,\varphi)t}a(\beta(\tau,\varphi)t;\tau,\varphi),\quad\beta(\tau,\varphi)<\frac{2\pi}{\tau},\quad\tau\ge0,\varphi\in S^{1}\label{eq:PRIMSETMW}
\end{equation}
is called primary set of MWs.} \emph{}\\
\emph{4. }\textbf{\emph{{[}Number of coexisting MWs{]}}}\emph{ With
the increasing of $\tau$, the number of MWs grows at least as $\tau^{2}$.
More specifically, suppose that (\ref{eq:eqper})--(\ref{eq:eqperbc})
has a regular MW solution $(a_{0}(\cdot),\omega_{0},\beta_{0})$,
$\beta_{0,}\omega_{0}\ne0$ for $(\tau_{0},\varphi_{0})$, which satisfies
the genericity condition 
\begin{equation}
\mathrm{rank}\left[\begin{array}{cc}
\partial_{\chi}T(\tau_{0},\varphi_{0}) & \partial_{\psi}T(\tau_{0},\varphi_{0})\\
\partial_{\chi}V(\tau_{0},\varphi_{0}) & \partial_{\psi}V(\tau_{0},\varphi_{0})
\end{array}\right]=2.\label{eq:gencond}
\end{equation}
where }$T(\tau,\varphi)=2\pi\beta^{-1}(\tau,\varphi)$, $V(\tau,\varphi)=T(\tau,\varphi)\omega(\tau,\varphi)$\emph{.
Then there exists a lower bound $\tau_{*}>0$ and a constant $c>0$
such that \eqref{eq:eqper}--\eqref{eq:eqperbc} has at least $c\tau^{2}$
MW solutions for all $\tau>\tau_{*}$ and all $\varphi\in[0,2\pi]$.
}\\
\emph{5. }\textbf{\emph{{[}Stability{]}}}\emph{ Stability of MW $e^{A\omega t}a(\beta t)$
is equivalent to the stability of the periodic solution $a(y)$ of
the DDE (\ref{eq:eqper}). In particular, if all multipliers except
two trivial ones have absolute values less than one, then the MW is
asymptotically exponentially stable. }\\
\emph{6. }\textbf{\emph{{[}Asymptotic stability for large delay{]}}}\emph{
For large delay, the asymptotic stability is determined by the asymptotic
continuous and strongly unstable point spectrum of Floquet exponents.
In particular, the MW $e^{A\omega t}a(\beta t)$ is exponentially
orbitally stable for large enough $\tau$ if all of the following
conditions hold:}

\emph{(i) the strongly unstable spectrum is empty;}

\emph{(ii) the Floquet exponent 0 has multiplicity 2 for sufficiently
large $\tau$, and}

\emph{(iii) except for the point $\mu=0$ for $\tilde{\varphi}$,
the asymptotic continouos spectrum is contained in $\left\{ z\in\mathbb{C}:\,\mathrm{Re}\, z<0\right\} .$}\\
\emph{The MW is exponentially unstable for all sufficiently large
$\tau$ if one of the following conditions holds:}

\emph{(j) the strongly unstable spectrum is non-empty, or }

\emph{(jj) a non-empty subset of the asymptotic continuous spectrum
has positive real part.}

Remarks and Examples.

\section{Proofs}

\subsection{Proof of Theorem 1 (CW)}

1. Statement 1 follows from Sec. \ref{sec:CW-MW}. 

2. Let $x(t;\tau_{0},\varphi_{0})=e^{A\omega(\tau_{0},\varphi_{0})t}x_{0}(\tau_{0},\varphi_{0})$
be a CW solution of (\ref{eq:main}), i.e. $\omega$ and $x_{0}$
satisfy (\ref{eq:eqforeq})--(\ref{eq:condx0}). Then for all parameter
values $(\tau,\varphi)$ such that (\ref{eq:phitaucw}) holds, we
have 
\[
e^{A(\varphi-\omega\tau)}=e^{A(\varphi_{0}-\omega_{0}\tau)}e^{A2\pi k}=e^{A(\varphi_{0}-\omega_{0}\tau)}
\]
and, hence, the same values of $\omega$ and $x_{0}$ satisfy (\ref{eq:eqforeq})
-- (\ref{eq:condx0}). 

3. The relation (\ref{eq:phitaucw}) implies that the same CW solution
$e^{A\omega(\tau_{0},\varphi_{0})t}x_{0}(\tau_{0},\varphi_{0})$ exists
for arbitrary $\tau=\tau_{1}\ge0$ provided $\varphi=\varphi_{1}$
with 
\[
\varphi_{1}=\varphi_{0}+\omega\left(\tau_{1}-\tau_{0}\right)\,\,\mathrm{mod}\,\,2\pi.
\]
In particular, it exists for $\tau=0$ and $\varphi=(\varphi_{0}-\omega\tau_{0})\,\mathrm{mod}\,2\pi$.
Hence, all CW solutions existing in (\ref{eq:main}) for arbitrary
values of $(\tau,\varphi)$, exist also for zero delay. Therefore,
we call the set of CW solutions for $\tau=0$ the \emph{primary set}.
The natural, possibly non-unique parametrization on this set is the
phase parameter $\varphi$. Since $f$ is smooth, the application
of implicit function theorem to system 
\[
f\left(X_{0}(\psi),e^{A\Psi}X_{0}(\Psi)\right)-A\Omega(\Psi)X_{0}(\Psi)=0\quad b^{T}X_{0}(\Psi)=0
\]
implies that there exist smooth branches of solutions $X_{0}^{(k)}(\Psi)$,
$\Omega^{(k)}(\Psi)$. 

4. In order to show that the number of CW solutions grows linearly
with delay, we consider any subset of the primary set of CW solutions
$x(t;\psi)=e^{A\Omega(\psi)t}X_{0}(\psi)$,\emph{ $\psi\in[\psi_{1},\psi_{2}]\in S^{1}$,
}for which $\Omega(\psi)$ is monotone. If the primary set corresponds
to the non-monotone function $\Omega(\psi)$, it can be decomposed
into a number of families with monotone $\Omega(\psi)$. 

\begin{figure}
\begin{centering}
\includegraphics[width=1\textwidth]{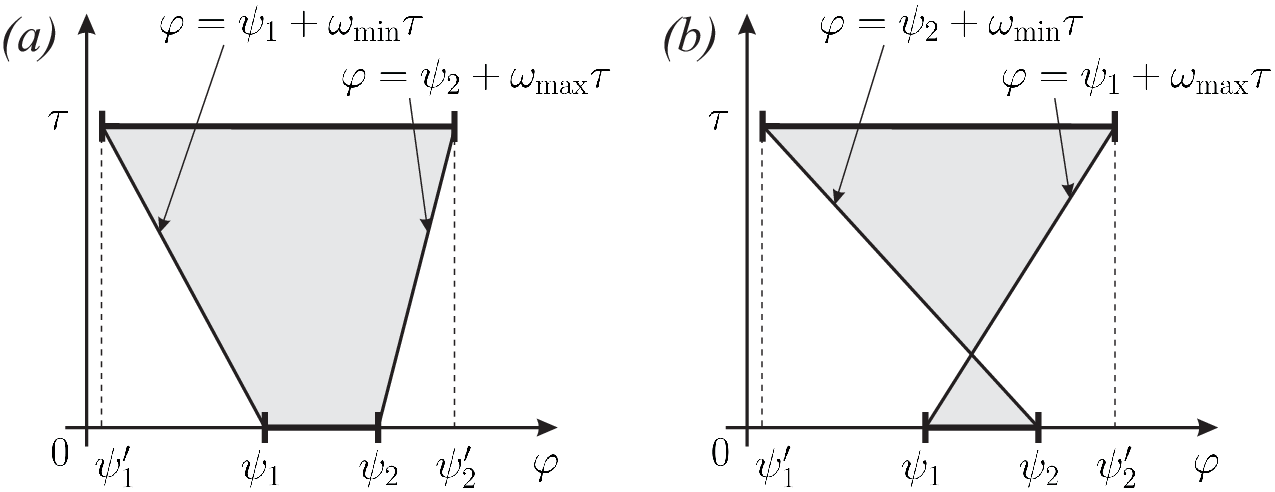}
\par\end{centering}

\caption{Reapearrance of CW solutions from some subset of the primary CW solutions
set. (a): case $\Omega_{\max}=\Omega(\psi_{2})$ and $\Omega_{\min}=\Omega(\psi_{1})$.
(b): case $\Omega_{\min}=\Omega(\psi_{2})$ and $\Omega_{\max}=\Omega(\psi_{1})$.\label{fig:CWREAP}}
\end{figure}

Accordingly to (\ref{eq:phitaucw}) the primary set reappears for
the parameters $(\tau,\varphi)$ satisfying
\begin{equation}
\varphi=\psi+\Omega(\psi)\tau.\label{eq:mapphi}
\end{equation}
Using the monotonicity of $\Omega(\psi)$, the interval $[\psi_{1},\psi_{2}]$
is mapped under the action of (\ref{eq:mapphi}) for some $\tau$
into the interval $[\psi_{1}',\psi_{2}']$ of the length $\Delta\psi'=\Delta\psi+\tau(\Omega_{\max}-\Omega_{\min})$,
where $\Delta\psi$ is either $\psi_{1}-\psi_{2}$ or $\psi_{2}-\psi_{1}$,
see Fig.~\ref{fig:CWREAP}. Since $\varphi$ is considered modulo
$2\pi$, the coexistence of multiple CWs is emerging due to the multiple
overlapping of the interval $\Delta\psi'$ over $2\pi$, i.e. the
number of CW solutions for delay $\tau$ is given by the integer value
of 
\[
\frac{\Delta\psi'}{2\pi}=\frac{\Delta\psi}{2\pi}+\frac{(\Omega_{\max}-\Omega_{\min})}{2\pi}\tau.
\]
The statement 4 follows from the fact that $\left|\Delta\psi\right|<2\pi$. 

5. In the corotating coordinates 
\[
x(t)=e^{A\omega t}y(t)
\]
 system (\ref{eq:main}) has the form 
\[
\frac{dy(t)}{dt}=f(y(t),e^{A\left(\varphi-\omega\tau\right)}y(t-\tau))-A\omega y(t)
\]
and the CW solution corresponds to the family of equilibria along
the group orbit 
\[
y(t)=x_{0}e^{i\gamma}.
\]
The characteristic equation for one of such equilibrium is independent
on $\gamma$ and reads 
\[
\det\left[\lambda I-D_{1}f(x_{0},e^{A\left(\varphi-\omega\tau\right)}x_{0})+A\omega-D_{2}f(x_{0},e^{A\left(\varphi-\omega\tau\right)}x_{0})e^{A\left(\varphi-\omega\tau\right)-\lambda\tau}\right]=0.
\]
Taking into account the relation $\varphi-\omega\tau=\psi\,\mathrm{mod}\,2\pi$
between $(\tau,\varphi)$ and the parameter $\psi$ on the primary
set, we obtain equation (\ref{eq:CWcheq}). Note that the equation
(\ref{eq:CWcheq}) has always one trivial eigenvalue $\lambda=0$,
which corresponds to the neutral direction along the group action.
In the original coordinates, this zero Lyapunov exponent corresponds
to the trivial zero Lyapunov exponent of the periodic solution. Hence,
the stability of the CW solution is determined by the remaining roots
of (\ref{eq:CWcheq}).

6. Asymptotic properties of roots of the characteristic equation for
general systems with one time delay for $\tau\to\infty$ have been
studied in \cite{Lichtner2011}. Statement 6 of the theorem follows
from the application of these results to (\ref{eq:CWcheq}).

\subsection{Proof of Theorem 2 (MW)}

1. Statement 1 follows from Sec. \ref{sub:CW}. 

2. Substituting (\ref{eq:taurule}) and (\ref{eq:phirule}) into (\ref{eq:eqper}),
we obtain the equation
\[
\beta\frac{da(y)}{dy}=-A\omega a(y)+f\left(a(y),e^{A\left(\varphi_{0}-\omega\tau_{0}\right)}a(y-\beta\tau_{0})\right),
\]
which, by assumption, has MW solution $e^{A\omega t}a(\beta t)$.
Hence, system (\ref{eq:main}) with parameters (\ref{eq:taurule})
-- (\ref{eq:phirule}) possesses the same MW solution.

3. Statement 3 follows directly from (\ref{eq:taurule}) -- (\ref{eq:phirule})
by taking minimal $k$, for which $\tau$ is positive. 

4. Denote the period $T_{0}=2\pi/\beta_{0}$, and the quantity $V_{0}=T_{0}\omega_{0}$.
The Implicit Function Theorem tells that for $(\chi,\psi)$ in a neighborhood
$\mathcal{P}$ of $(\tau_{0},\phi_{0})$ the BVP \eqref{eq:eqper}--\eqref{eq:eqperbc}
defines locally a surface of solutions, parametrized by $\chi$ and
$\psi$: $(a(\chi,\psi)(\cdot),\beta(\chi,\psi),\omega(\chi,\psi))$.
For these solutions in the vicinity $\mathcal{P}$ we also have the
quantities 
\[
\begin{aligned}T(\chi,\psi) & =2\pi/\beta(\chi,\psi) &  & \mbox{(the period),}\\
V(\chi,\psi) & =T(\chi,\psi)\omega(\chi,\psi) &  & \mbox{(the phase after one period).}
\end{aligned}
\]
The starting point of the proof is the relation (\ref{eq:taurule})-(\ref{eq:phirule}).
If $\chi$ and $\psi$ are in the neighborhood $\mathcal{P}$ of $(\tau_{0},\varphi_{0})$,
and we can find integers $k$ and $\ell$ such that 
\begin{equation}
\begin{split}T(\chi,\psi)+\frac{\chi}{k} & =\frac{\tau}{k}\\
V(\chi,\psi)+\frac{\psi}{k} & =\frac{\varphi}{k}+\frac{2\pi\ell}{k}
\end{split}
\label{eq:recursion}
\end{equation}
then the solution $(a(\chi,\psi)(\cdot),\omega(\chi,\psi),\beta(\chi,\psi))$
is also a MW solution of \eqref{eq:eqper}--\eqref{eq:eqperbc} for
the parameter values $(\tau,\varphi)$ appearing in the right-hand
side of \eqref{eq:recursion}.

We observe that the primary solution $(a_{0}(\cdot),\omega_{0},\beta_{0})$
at $(\tau_{0},\varphi_{0})$ satisfies the following (trivial) relation:
\begin{equation}
\begin{split}T(\tau_{0},\varphi_{0}) & =T_{0}\mbox{,}\\
V(\tau_{0},\varphi_{0}) & =V_{0}\mbox{.}
\end{split}
\label{eq:trivial}
\end{equation}
 Note we can assume that $V_{0}\in[\pi,3\pi]$ without loss of generality
because one can add arbitrary integer multiples of $\beta_{0}$ to
$\omega_{0}$. Indeed, changing the definition of $a_{0}$ accordingly
to $\exp(-A\beta_{0}t)a_{0}(\beta_{0}t)$ adds integer multiples of
$2\pi$ to $V_{0}$. Genericity condition \eqref{eq:gencond} and
the relation \eqref{eq:trivial} ensure that the system 
\begin{equation}
\begin{split}T(\chi,\psi)+\epsilon\chi & =T_{0}+\delta\mbox{,}\\
V(\chi,\psi)+\epsilon\psi & =V_{0}+\gamma
\end{split}
\label{eq:perturb}
\end{equation}
 has a locally unique solution $(\chi,\psi)\in\mathcal{P}$ for all
sufficiently small $\epsilon$, $\delta$ and $\gamma$, that is,
$\epsilon$, $\delta$ and $\gamma$ satisfying 
\[
|\epsilon|<\epsilon_{0}\mbox{,\quad}|\delta|<\delta_{0}\mbox{,\quad}|\gamma|<\gamma_{0}
\]
for some $\epsilon_{0}>0$, $\delta_{0}>0$ and $\gamma_{0}\in(0,2\pi)$.
The remaining step is that we have to count for any given $\tau$
and $\varphi$ how many integer pairs $\left(k,\ell\right)$ satisfy
the relations 
\begin{align}
1/\epsilon_{0} & <k\mbox{,}\label{eq:epsbound}\\
\delta_{0} & >\left|\frac{\tau}{k}-T_{0}\right|\mbox{,}\label{eq:kbound-1}\\
\gamma_{0} & >\left|\frac{\varphi}{k}-V_{0}+\frac{2\pi\ell}{k}\right|\mbox{.}\label{eq:lbound}
\end{align}
For each integer pair $(k,\ell)$ satisfying \eqref{eq:epsbound}--\eqref{eq:lbound}
the recurrence \eqref{eq:recursion} has a locally unique solution
$(\chi,\psi)$, corresponding to a modulated wave.

A small side calculation (follows below) gives the following result:
if 
\begin{equation}
\tau>\tau_{*}:=\max\left\{ \frac{2T_{0}}{\epsilon_{0}},\frac{8\pi T_{0}}{\gamma_{0}}\right\} \label{eq:taubound}
\end{equation}
 then we can find at least 
\begin{align}
 & \mathrm{floor}\left(\frac{2r}{T_{0}}\tau\right)\mbox{\quad integers \ensuremath{k}, and}\label{eq:kcount}\\
 & \mathrm{floor}\left(\frac{\gamma_{0}-6\pi r}{2\pi T_{0}}\tau\right)\mbox{\quad integers \ensuremath{\ell},}\label{eq:lcount}
\end{align}
satisfying the conditions \eqref{eq:epsbound}--\eqref{eq:lbound}.
In \eqref{eq:kcount}--\eqref{eq:lcount} the quantity $r$ is chosen
such that it satisfies 
\begin{equation}
0<r<\min\left\{ 1,\frac{\delta_{0}}{T_{0}},\frac{\gamma_{0}}{6\pi}\right\} \mbox{,}\label{eq:rdef}
\end{equation}
such that the pre-factors of $\tau$ in both counts \eqref{eq:kcount}
and \eqref{eq:lcount} are positive.

Side calculation to show \eqref{eq:kcount}--\eqref{eq:lcount}

First, let us consider integers $k$ that satisfy 
\begin{equation}
\frac{\tau}{T_{0}(1+r)}\leq k\leq\frac{\tau}{T_{0}(1-r)}\mbox{.}\label{eq:kineq}
\end{equation}
 Since $r<1$ the right bound is indeed positive. How many integer
are these? The difference between upper and lower bound is larger
then $2r\tau/T_{0}$, which is the number we claim in \eqref{eq:kcount}
to satisfy the bounds. If $k$ satisfies \eqref{eq:kineq} then $\tau/k-T_{0}$
satisfies 
\[
-T_{0}r\leq\frac{\tau}{k}-T_{0}\leq T_{0}r\mbox{,}
\]
 and, because we have chosen $r<\delta_{0}/T_{0}$, 
\[
\left|\frac{\tau}{k}-T_{0}\right|<\delta_{0}\mbox{,}
\]
which is what requirement \eqref{eq:kbound-1} needs. Furthermore,
since $r<1$ and $\tau>2T_{0}/\epsilon_{0}$, the lower bound in \eqref{eq:kineq}
is greater than $1/\epsilon_{0}$, such that $k$ also satisifes requirement
\eqref{eq:epsbound}.

We now try to figure out, which integers $\ell$ satisfy requirement
\eqref{eq:lbound}. First, we observe that the term $\phi/k$ is smaller
in modulus than $\gamma_{0}/2$ for all $\phi\in[0,2\pi]$: we have
that $k>\tau/(2T_{0})$ due to \eqref{eq:kineq}, hence, 
\[
|\phi/k|<4\pi T_{0}/\tau\leq\gamma_{0}/2\mbox{,}
\]
 because we chose $\tau>\tau_{*}\geq8\pi T_{0}/\gamma_{0}$ in \eqref{eq:taubound}.
Consequently, if we find integers $\ell$ satisfying 
\begin{equation}
|2\pi\ell/k-V_{0}|<\gamma_{0}/2\label{eq:lineq1}
\end{equation}
 for all $k\in[\tau/(T_{0}(1+r)),\tau/(T_{0}(1-r))]$, these integers
$\ell$ will also satisfy the bound \eqref{eq:lbound} for all $k\in[\tau/(T_{0}(1+r)),\tau/(T_{0}(1-r))]$.
The inequality \eqref{eq:lineq1} is equivalent to 
\begin{equation}
V_{0}-\gamma_{0}/2\leq2\pi\ell/k\leq V_{0}+\gamma_{0}/2\mbox{,}\label{eq:lineq2}
\end{equation}
 where the lower boundary is positive ($V_{0}\geq\pi$ and $\gamma<2\pi$).
The positive integer $k$ is bounded by \eqref{eq:kineq}, which means
that \eqref{eq:lineq2} would follow from 
\begin{equation}
\frac{V_{0}-\gamma_{0}/2}{2\pi T_{0}}(1+r)\tau\leq\ell\leq\frac{V_{0}+\gamma_{0}/2}{2\pi T_{0}}(1-r)\tau\mbox{.}\label{eq:lineq3}
\end{equation}
 How many integers $\ell$ fit between the bounds in \eqref{eq:lineq3}?
The difference between upper and lower bound is 
\begin{equation}
\frac{\gamma_{0}-2V_{0}r}{2\pi T_{0}}\tau\mbox{.}\label{eq:lcountrough}
\end{equation}
 Since $V_{0}\in[\pi,3\pi]$ and $r<\gamma_{0}/(6\pi)$ the pre-factor
of $\tau$ in expression \eqref{eq:lcountrough} is positive. Replacing
$V_{0}$ by its upper bound $3\pi$ makes \eqref{eq:lcountrough}
smaller and equal to the claimed number \eqref{eq:lcount} (but it
is still a positve multiple of $\tau$). \hfill{}(\emph{End of side
calculation})

Consequently, if we choose $\tau$ larger than the $\tau_{*}$ given
in \eqref{eq:taubound} we find 
\begin{equation}
N(\tau):=\mathrm{floor}\left(\frac{2r}{T_{0}}\tau\right)\times\mathrm{floor}\left(\frac{\gamma_{0}-6\pi r}{2\pi T_{0}}\tau\right)\label{eq:klcount}
\end{equation}
pairs of integers $\left(k,\ell\right)$ for which the recurrence
\eqref{eq:recursion} has a solution $(\chi,\psi)$ in the neighborhood
$\mathcal{P}$, and, hence, the BVP \eqref{eq:eqper}--\eqref{eq:eqperbc}
has a modulated wave. If necessary, we increase $\tau_{*}$ such that
$N(\tau_{*})>0$. Then $N(\tau)/\tau^{2}$ is uniformly positive for
all $\tau>\tau_{*}$, such that we can choose the constant $c$ in
the claim of the lemma as $c:=\inf_{\tau\geq\tau_{*}}N(\tau)/\tau^{2}>0$.

At last, let us show that every different pair $(k,\ell)$ leads to
a different MW. For this, it enough to show that every different pair
$(k,\ell)$ corresponds to a different pair $(\chi,\psi)$, as a solution
of (\ref{eq:recursion}). Indeed, in a small neighborhood of $(\tau_{0},\varphi_{0})$,
due to the nondegeneracy condition \eqref{eq:gencond}, we have 
\[
\det\left[\begin{array}{cc}
-\frac{1}{\beta^{2}}\beta_{\chi} & \frac{1}{\beta^{2}}\beta_{\psi}\\
-\frac{1}{\beta^{2}}\beta_{\chi}\omega+\beta\omega_{\chi} & -\frac{1}{\beta^{2}}\beta_{\psi}\omega+\beta\omega_{\psi}
\end{array}\right]=\det\left[\begin{array}{cc}
-\frac{1}{\beta^{2}}\beta_{\chi} & \frac{1}{\beta^{2}}\beta_{\psi}\\
\beta\omega_{\chi} & \beta\omega_{\psi}
\end{array}\right]\ne0,
\]
which implies 
\[
\det\left[\begin{array}{cc}
\beta_{\chi} & \beta_{\psi}\\
\omega_{\chi} & \omega_{\psi}
\end{array}\right]\ne0.
\]
Hence, the mapping $(\chi,\psi)\to(\beta,\omega)$ is homeomorphism
and maps different pairs of $(\chi,\psi)$ to different pairs $(\beta,\omega)$.
Now let us assume that two pairs $(k_{1},\ell_{1})$ and $(k_{2},\ell_{2})$
solve (\ref{eq:recursion}) with the same pair $(\chi,\psi)$. The
first equation in (\ref{eq:recursion}) implies that $\chi\ne\tau$
and 
\[
\frac{\tau-\chi}{k_{2}}=\frac{\tau-\chi}{k_{1}}
\]
leading to $k_{1}=k_{2}$. Similarly, the second equation in (\ref{eq:recursion})
implies $\ell_{1}=\ell_{2}$. Hence, we have shown that every different
pair $(k,\ell)$ leads to a different MW.

5. The variational equation to (\ref{eq:eqper}) has the form 
\[
\beta\frac{dv(y)}{dy}=\left[-A\omega+D_{1}f\left(a(y),e^{A\left(\varphi-\omega\tau\right)}a(y-\beta\tau)\right)\right]v(y)
\]
\begin{equation}
+D_{2}f\left(a(y),e^{A\left(\varphi-\omega\tau\right)}a(y-\beta\tau)\right)e^{A\left(\varphi-\omega\tau\right)}v(y-\beta\tau).\label{eq:var}
\end{equation}
Two trivial multipliers correspond to the following solutions 
\begin{equation}
\frac{da(y)}{dy},\quad Aa(y)\label{eq:nontr}
\end{equation}
of the variational equation.

6. Application of the result of \cite{Sieber2013}.

\bibliographystyle{plain}

\end{document}